\documentclass[12pt,a4paper]{article}
\usepackage{latexsym}
\usepackage{amsfonts}

\newcommand{\danger}[1]{\textbf{#1}}

\input{epsf}             

\usepackage[dvips]{graphicx}

\begin{document}

\title{\danger{BTZ Black Hole Entropy: A spin foam model description}}
\author{\centerline{\danger{J. Manuel Garc\'\i a-Islas \footnote{
e-mail: jmgislas@leibniz.iimas.unam.mx}}}  \\
Instituto de Investigaciones en Matem\'aticas Aplicadas y en Sistemas \\ 
Universidad Nacional Aut\'onoma de M\'exico, UNAM \\
A. Postal 20-726, 01000, M\'exico DF, M\'exico\\}

\maketitle

\danger{Abstract}. We present a microscopical explanation of the
entropy of the BTZ black hole using discrete spin foam models of
quantum gravity. The entropy of a black hole is given in geometrical
terms which lead us to think that its statistical description must
be given in terms of a quantum geometry. In this paper we present it
in terms of spin foam geometrical observables at the horizon of the
black hole.

\section{\bf{Introduction}}

Since its intoduction in \cite{b} and \cite{h}, black hole entropy
has intrigued the physical community and attempts for a statistical
explanation of it have appeared since then in the literature.
Quantum gravity is believed to play a major role in the explanation
of black hole entropy and different approaches have studied the
problem, from string theory to loop quantum gravity. Black hole
solutions in three dimensions were discovered for the first time by
Ba\~nados,Teitelboim and Zanelli \cite{btz}, and then the entropy
problem has also been studied for this case in many different papers
by different approaches \cite{ct}, \cite{c}, \cite{s} \cite{kk},
just to mention some references.

In this paper we study the case of the BTZ black hole using spin
foam models. Spin foam models are the covariant description of loop
quantum gravity. The description we follow is given in terms of
geometrical observables at the horizon of the black hole. These
observables have been introduced and studied in \cite{gi},
\cite{bgm}. We work with the Euclidean BTZ black hole. The Euclidean
version is known to be topologically a solid torus where the core of
the torus is the horizon. The idea is to think of the three
dimensional black hole manifold as a spin foam. This lead us to
triangulate the solid torus where in addition the horizon is
considered as an observable. This can be thought as a new example of
spin foam models with observables, in the same context of \cite{gi},
\cite{bgm}. The black hole entropy is shown to be related to the
logarithm of the expectation value of the horizon observable. The
main contribution is shown to be proportional to the horizon length
as expected. In this way we are saying that the statistical
description of the black hole entropy is given in terms of some
geometrical observables in the horizon.

This description is of course a discrete one which comes from spin
foam models. A description of the black hole entropy in terms of the
Ponzano-Regge model was studied in \cite{skg}.

We divide this paper as follows. In section 2 we very briefly
describe the three dimensional Euclidean BTZ black hole. In section
3 we describe 3-dimensional spin foam models in a general sense. In
section 4 we review the notion of observables introduced in
\cite{gi} and \cite{bgm}. Section 5 is the main and most important
part of the paper where we develop a study of the entropy of BTZ
black hole as a spin foam model description.

\section{The BTZ black hole}

In this section we describe the Euclidean BTZ black hole which was
introduced for the first time in \cite{btz}. The three dimensional
Euclidean solution to empty Einstein equations of general relativity
with negative cosmological constant is given by the metric

\begin{equation}
ds^{2}= \bigg(\frac{R^{2}}{\ell^{2}}-M\bigg) d\tau^{2} +
\bigg(\frac{R^{2}}{\ell^{2}}-M\bigg)^{-1} dR^{2} +r^{2} d\phi^{2}
\end{equation}
In \cite{c}, it is shown that by a change of coordinates, the
solution can be written in the form

\begin{equation}
ds^{2}= \frac{\ell}{z^{2}}(dx^{2} + dy^{2} + dz^{2})
\end{equation}
for $z>0$. Immediately it can be recognised as the metric of the
hyperbolic space $H^3$. Then after some isometric identifications
the BTZ solution is in fact given by a fundamental region of the
hyperbolic space. This region is a solid torus where the core of the
torus is the black hole horizon $R= \ell\sqrt{M}$, and the rest of
the torus is the outside of the black hole $R> \ell\sqrt{M}$.

Observe that the solution implies that the black hole has a constant
negative curvature which differs from four dimensional gravity where
black hole solutions have variable curvature which grows indefinitely as we
approach the singularity.

According to Bekenstein-Hawking formula the leading term of the
entropy of a black hole is given by

\begin{equation}
S= \frac{A}{4}
\end{equation}
where $A$ is the black hole horizon area. In the case of the BTZ
black hole, as its horizon is a one dimensional circle this formula
is obviously given by analogy

\begin{equation}
S= \frac{L}{4}
\end{equation}
where $L$ is the black hole horizon length. The entropy is believed
to be related to the logarithm of the number of microstates. In this
paper we propose a way to define the microstates of the black hole.

As the entropy of the black hole is related to a geometric property
of the black hole, in this case, the length of the horizon, it is
quite natural to think that its microstate description should come
from a kind of quantum geometric property. This is the approach we
follow and it is done in section 5.

\section{Three Dimensional Spin Foam Model}

In this section we describe three dimensional spin foam models in
general, where the Turaev-Viro model \cite{tv} is included. We
define a three dimensional spin foam model in the following way.
Consider a three dimensional space-time manifold $M$ which is
assumed to be compact and oriented. Then we triangulate it
$\triangle$ where the triangulation is composed by vertices, edges,
triangles and tetrahedra. It is more common to define spin foam
models in the dual complex of the triangulation, however we stick to
the triangulation. Consider a set of indices $L= \{
0,1/2,1,...(r-2)/2 \}$, where $r\geq3$. Then we define a state as a
function from the set of edges $S: \{edges\} \longrightarrow L$. A
state is called admissible if at each face of the triangulation, the
labels $(i,j,k)$ of the corresponding edges satisfy the following
identities:

\begin{eqnarray}
0 \leq i , j , k \leq \frac{r-2}{2} \nonumber \\
i \leq j+k , \  j \leq i+k , \  k \leq i+j \nonumber \\
i+j+k \equiv mod \ 1 \nonumber \\
i+j+k \leq (r-2)
\end{eqnarray}
The spin foam partition function is then given by

\begin{equation}
Z(M)=  \sum_{S} \prod_{edges} dim_{q}(j) \prod_{tetrahadra} \{ 6j \}
\end{equation}
where the sum is carried over the set of all admissible states
$S$.\footnote{In order to obtain a topological invariant of the
manifold the sum above is multiplied by a factor $N^{-v}$ given by
$N= \sum (dim_{q}(j))^{2}$, where $v$ is the number of vertices of
the triangulation. We do not use this factor in this paper} The
weight $\{ 6j \}$ is the $6j$ symbol associated to the six labels of
each tetrahedron and the quantum dimension is given by the quantum
number
$dim_{q}(j)=[2j+1]_{q}=(-1)^{2j}(\sin(\pi(2j+1)/r))/(sin(\pi/r)$.
When the three dimensional manifold has a boundary as it is in our
case, we have a similar state sum as above with a slight variation
where all the states on the boundary are kept fixed, and each edge
in the triangulation of the boundary gives a contribution of
$dim_{q}(j)^{1/2}$.

\section{The Observables}

In \cite{gi} and \cite{bgm} a notion of observables for the
Turaev-Viro spin foam model is introduced. The observables are
defined as follows; given a triangulation $\triangle$ of our three
dimensional space-time manifold $M$, we consider any subset we want
of edges of the triangulation. We denote this subset by
$\mathcal{O}$. This subset can be topologically a tree, graph with
circuits, knot or any combination of these examples. We now colour
our triangulation with spins from the subset $L= \{
0,1/2,1,...(r-2)/2 \}$ and consider the following partition function

\begin{equation}
Z(M, \mathcal{O}(j_{1}, j_{2},...,j_{k})) = \sum_{S \mid
\mathcal{O}} \prod_{edges}dim_{q}(j) \prod_{tetrahedra} \{ 6j \}
\end{equation}
which is similar to the state sum formula (6) with the only
difference that now we are not summing over the spins which label
the graph observable $\mathcal{O}$. That is, the spins labelling the
graph observable $\mathcal{O}$ are kept fixed. The above partition
function is then a function of the spins $j_{1}, j_{2},...,j_{k}$
which label the observable. The expectation value of the observable
$\mathcal{O}$ is then given by

\begin{equation}
W(M, \mathcal{O}(j_{1}, j_{2},...,j_{k}))=
\frac{Z(M,\mathcal{O}(j_{1}, j_{2},...,j_{k}))}{Z(M)}
\end{equation}
This expectation value is well defined when $Z(M) \neq 0$.

Some examples of observables and of their expectation value are
given in \cite{gi}. For instance if the observable $\mathcal{O}$
consists of a single edge $e$, we have that its expectation value
does not depend on the manifold $M$ in which it lives and the value
is given by

\begin{equation}
W(M, \mathcal{O}(j))= dim_{q}(j)^{2}
\end{equation}
If the observable $\mathcal{O}$ consists of a triangle whose edges
are labelled by $i$, $j$ and $c$, we have that its expectation value
is again independent of the manifold $M$ in which it lives and its
value is given by

\begin{equation}
W(M, \mathcal{O}(i,j,c))= N_{i,j,c} \dim_{q}(i) \dim_{q}(j)
\dim_{q}(c)
\end{equation}
where $N_{i,j,c}$ is the dimension of the space of intertwiners,
i.e. equal to 1 if the spins are admissible and 0 otherwise. Another
example will be given in the next section and indeed it is the most
important for us in this paper.

\section{Black hole entropy}

In this section we apply the notion of observables, reviewed in the
previous section, to the case of the BTZ black hole. This is a new
example of observables in three dimensional quantum gravity as
introduced in the previous section. We will think of the horizon as
an observable in the three dimensional space-time given by the
Euclidean BTZ black hole.

Our spin foam model description of the BTZ black hole will be in the following
way. We think of the black hole space-time as a 
discrete  two complex or a triangulated manifold.
Therefore we start by triangulating the Euclidean black
hole. There are an infinite number of triangulations of the solid
torus which may or may not contain any interior edges. We will
consider triangulations of the solid torus which contain interior
edges, as we want the core of the torus(horizon) be formed by edges.

\begin{figure}[h]
\begin{center}
\includegraphics[width=0.3\textwidth]{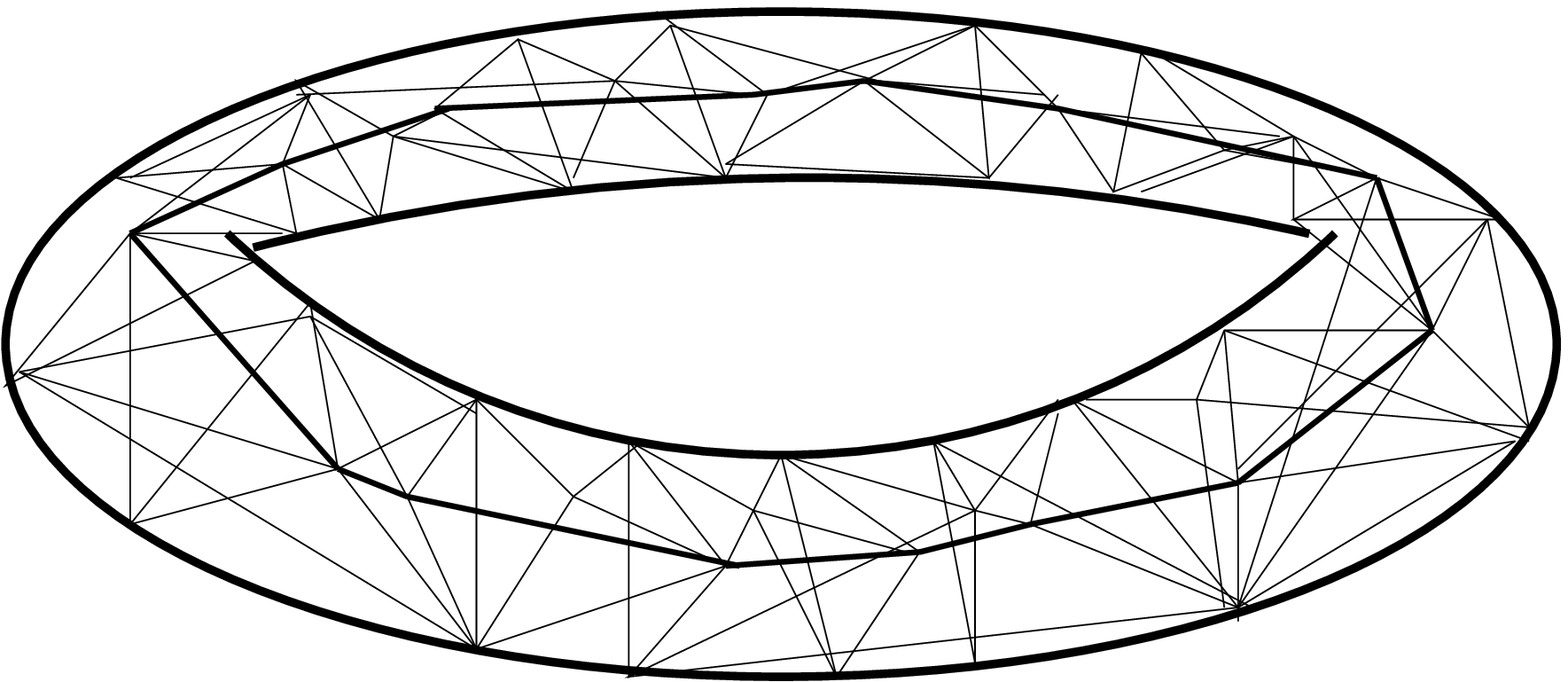}
\caption{Triangulation of the BTZ Euclidean black hole with interior
edges}
\end{center}
\end{figure}
We now think of the horizon as an observable and consider the
expectation value of the observable, that is, if we denote the
partition function of the solid torus with its core as a graph
observable by $Z(T^2,\mathcal{O})$ we want to calculate

\begin{equation}
W(T^2, \mathcal{O})= \frac{Z(T^2, \mathcal{O})}{Z(T^2)}
\end{equation}
It is important to note that our expectation value given by formula (11)
will depend on the triangulation we take. 
This is obvious since in our definition of
the three dimensional spin foam model in section $3$ we do not have the
common regularisation factor used in the Turaev-Viro model.  

But dependence in the triangulation is what we want. Our explanation for this
is as follows. Our expectation value will depend not only in the triangulation we take but 
in the spins which label the horizon. In \cite{jb} it is explained how a labelled edge  
can be interpreted for instance as giving length to the corresponding edge. This is the idea
we take so that the expectation value of our observable  is related to the length
of our horizon. Length is a geometrical entity and not topological. This is why we want
our expectation value to be triangulation dependent.  Moreover if different triangulations are used 
to define $Z(T^2, \mathcal{O})$ and $Z(T^2)$ it could be thought that we still have an ambiguity since 
both functions do not include the regularisation factor used to make it topological. However
for the particular triangulations we consider in this paper there is no ambiguity in the definition of formula
$(11)$ as the triangulation which defines $Z(T^2, \mathcal{O})$ will be related to the triangulation which defines $Z(T^2)$ by simple subdivisions.

We now carry on with our computation.
The partition function of the solid torus without
the observable $\mathcal{O}$, that is $Z(T^2)$ is given as follows.
The solid tours can be triangulated with $n$ tetrahedra without
interior edges and vertices. This means that it can be triangulated
in a way in which only exterior edges exist. 
Let us label the exterior edges by $\widehat{j}$.
The partition function is therefore given by

\begin{equation}
Z(T^2)= \prod_{edges} dim_{q}(\widehat{j})^{1/2}
\prod_{tetrahedra}^{n} \{ 6\widehat{j} \}
\end{equation}
The upper $n$ in the product symbol denotes that we have $n$
tetrahedra and therefore a product of $n$ $6j$ symbols. We now show
how to calculate the value of $Z(T^2, \mathcal{O})$. With the same
solid torus we used for formula (12), that is, with the same number
$n$ of tetrahedra and the same labels at all of the exterior edges,
apply Pachner move $1-4$ to all of the tetrahedra, fig[2].

\begin{figure}[h]
\begin{center}
\includegraphics[width=0.3\textwidth]{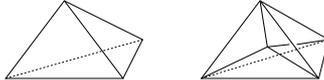}
\caption{Pachner move 1-4}
\end{center}
\end{figure}
We can alternatively think that we are subdividing each tetrahedron of the triangulated solid
tours. We are applying Pachner move just as a procedure in order to have interior edges which will
discretize our horizon.  We are just considering a particular and specific triangulation.  
We also have some horizon vertices which touch the boundary of the solid torus. This is not
a problem since they are isolated.  

In this way we have a triangulation of the solid torus with $4n$
interior tetrahedra. We also have interior edges which go around the
torus, and can think of any of these set of edges which form a
closed path, as our horizon. We now take the partition function with
the horizon as an observable. Observe that for each tetrahedron
which was subdivided using Pachner move $1-4$, two interior edges
belong to the horizon and the remaining ones are summed over all
possible states in the partition function. It is not difficult to
realise that our partition function $Z(T^2, \mathcal{O})$ is given
by

\begin{equation}
Z(T^2, \mathcal{O})= \prod_{exterior edges}
dim_{q}(\widehat{j})^{1/2} \sum_{S \mid \mathcal{O}} \prod_{interior
edges} dim_{q}(j) \prod_{tetrahedra} \{ 6j \}
\end{equation}
Observe that the $6j$ symbols which appear in the above formula are
not the same as the ones of formula (12). In formula (13) we have
$4n$ tetrahedra. However the product of the exterior edges of both
formulas (12) and (13) are the same, as we have not changed the
labels of the exterior edges of the torus. 
It is easier to realize how the calculation goes if we think of the partition function locally.
And this can be done in our case since each exterior tetrahedron was subdivided in
$4$ tetrahedra with interior edges.
Note that the partition
function sum $Z(T^2, \mathcal{O})$ reads locally

\begin{equation}
dim_{q}(\widehat{j_1})^{1/2} \cdots dim_{q}(\widehat{j_6})^{1/2}
\sum_{k,l} dim_{q}(i) dim_{q}(k) dim_{q}(l) dim_{q}(j) \times \]
\[ \qquad \qquad \times
\left(
\begin{array}{ccc}
\widehat{j_1}&\widehat{j_2}&\widehat{j_3}\\
 j&i&l\\
  \end{array}
\right)_{q}
\left(
\begin{array}{ccc}
\widehat{j_6}&\widehat{j_5}&\widehat{j_1}\\
 i&l&k\\
  \end{array}
\right)_{q}
\left(
\begin{array}{ccc}
\widehat{j_4}&\widehat{j_2}&\widehat{j_6}\\
 l&k&j\\
  \end{array}
\right)_{q}
\left(
\begin{array}{ccc}
\widehat{j_3}&\widehat{j_5}&\widehat{j_4}\\
 k&j&i\\
  \end{array}
\right)_{q}
\end{equation}
where $i,j$ are labels of edges which belong to the discrete
horizon. Observe that these labels $i,j$ are fixed. Summing over $k$
and $l$ and using the Biedenharn-Elliot identity, orthogonality and
symmetry properties \cite{kl} we get

\begin{equation}
dim_{q}(\widehat{j_1})^{1/2} \cdots
dim_{q}(\widehat{j_6})^{1/2}\left(
\begin{array}{ccc}
\widehat{j_1}&\widehat{j_2}&\widehat{j_3}\\
 \widehat{j_4}&\widehat{j_5}&\widehat{j_6}\\
  \end{array}
  \right)_{q}
\frac{N_{i,j,\widehat{j_3}}}{dim_{q}(\widehat{j_3})} dim_{q}(i)
dim_{q}(j)
\end{equation}
This calculation has reduced four $6j$ symbols in the interior of
the solid torus, to only one tetrahedron which edges belong to the
boundary. From this observation it it easy to deduce that the
expectation value of our observable is given as follows: the horizon
is triangulated with $2n$ edges.  Label the horizon edges by
$i_{1},j_{1} ,\cdots ,i_{n},j_{n}$. Each pair of edges $i_m,j_m$
belong to a triangle. The remaining edges which belong to these $n$
triangles belong to the boundary of the solid torus. Label these
edges by $\widehat{j_1}, \cdots, \widehat{j_n}$. The expectation
value is therefore given by

\begin{equation}
W(T^2, \mathcal{O})= \prod_{m}
\frac{N_{i_m,j_{m},\widehat{j_m}}}{dim_{q}(\widehat{j_m})}
dim_{q}(i_m) dim_{q}(j_m)
\end{equation}
where $m=1,\cdots,n$. Recall that the factor
$N_{i_m,j_{m},\widehat{j_m}}$ is zero if the states are non
admissible and $1$ if states are admissible. We are considering
admissible states since they will lead us to a non zero calculation.
By relabelling edges of the horizon, the entropy is then given by

\begin{equation}
S= \sum_{m=1}^{2n} \log(dim_{q}(j_m)) -
\sum_{k=1}^{n}\log(dim_{q}(\widehat{j_k)})
\end{equation}
Recall that the quantum dimensions are given by the quantum numbers
$dim_{q}(j)=[2j+1]_{q}$. Note a similarity between our entropy
formula (17) and the study of the entanglement entropy in \cite{d}.

We see that our formula (17) is triangulation dependent and it also depends
on the spins which label the horizon and some of the spins which label
boundary edges. 

We could have a different labeling of the same triangulation and different triangulations.
First of all, for a fixed triangulation the question is whether we should sum over all admissible
configurations to account for the entropy. 
We propose that this should be the case although in this paper we just consider a simpler
configuration.

 Let us analyse our formula (17) of the entropy of the BTZ black hole
in order to see how it relates to the length of the horizon.

It is important to mention some important facts. The labels of the
edges of the horizon by spins $j$ are interpreted in the spin foam
model as giving a discrete length. Following the arguments given in
\cite{jb}, in the present case a spin $j$, can be interpreted as
having length $j + \frac{1}{2}$.

The horizon is discrete and formed by edges with spins $j_1,\cdots
j_{2n}$. We have a constraint, as we want that the sum of all the
discrete lengths of the horizon be $L$. The length of a circle is
given by $2 \pi R$ where $R$ is the radius of the circle. This is

\begin{equation}
j_{1}+ \cdots +j_{2n}+ \frac{2n}{2}=  2 \pi R
\end{equation}
From formula (17) it is obvious that the largest contribution to the
entropy is given when the second sum vanishes. 
This implies that when we consider the main contribution to the entropy
it only matters what happens at the horizon. 
The microstates live at the horizon. 
This means that
we should only care about the number of edges of the horizon
and the spins which label them.
This is given when
all of the spins $\widehat{j_k}$ are equal to zero, that is, when
all these edges are removed from the triangulation. 
In this case it
implies by (5) that the spins are equal in pairs $j_1=j_2,...,j_{2n-1}=j_{2n}$.
It can be observed now that the largest contribution 
is given when the spins of the horizon are all
equal, which turns formula (18) into

\begin{equation}
2nj+ \frac{2n}{2}= 2 \pi R
\end{equation}
and formula (17) into

\begin{equation}
S= 2n \log(dim_{q}(j))
\end{equation}
and a substitution of formula (19) into (20) we have

\begin{equation}
S = \frac{2 \pi R}{j+\frac{1}{2}} \log (dim_{q}j)
\end{equation}
which shows that the main contribution to the entropy is
proportional to the length of the horizon. 
Call $L=2 \pi R$.
The contribution is
bigger when the spin $j=1$, and when $r$ is large we have
$dim_{q}(j)\rightarrow(2j+1)$, which shows that formula (21) is
dominated by

\begin{equation}
S \simeq  \frac{2}{3} \log (3) L
\end{equation}
which shows that the entropy is proportional to the length of the horizon.
The main contribution to the entropy is given by the previous
formula. however if we want to compute a contribution from all the
possible states, the problem turns into a combinatorial one as we
now propose. According to our approach to the entropy of BTZ black
hole, the contributions come from the edges which form the horizon,
and from the ones at the boundary which they form triangles with.
All of these edges are fixed and chosen from the very beginning. We
have $2n$ edges at the horizon and $n$ edges at the boundary. This
means that we have a set of $n$ triangles which we label
$\{\{i_1,j_1,\widehat{j_1}\}, \cdots ,\{i_n,j_n,\widehat{j_n}\}\}$.
Each labelled triangle should satisfy the admissible conditions of
formula (5). This means that computing the entropy requires also
solving the combinatorial problem of counting all of the admissible
configurations for every finite triangulation.
In other words, what we are saying is that when we consider a fixed macrostate
given by a fixed length, the microstates which account for that entropy are the number
of different triangulations and spins which satisfy equality $(18)$.  

This resembles very well the approach done in Loop Quantum Gravity; however
we do not have the famous Immirzi parameter. In this description we should therefore 
be able to get  the famous $1/4$ factor 
somehow. For the moment we do not know the way to solve this problem.

\bigskip

\danger{Acknowledgement}: I want to thank Alejandro Corichi for his
suggestion on this problem. I also thank John Barrett and Steven Carlip for their
e-mail correspondence which led to some improvements on this paper.


\begin{thebibliography}{99}

\bibitem{b} J.D. Bekenstein, Black Holes and Entropy, Phys. Rev D. \danger{7} (1973)
         2333-2346


\bibitem{h} S.W. Hawking, Particle Creation by Black Holes, Commun. Math. Phys \danger{43} (1975)
           199-220


\bibitem{btz} M. Ba\~nados, C. Teitelboim, J. Zanelli, Black Hole in Three-Dimensional
              Spacetime, Phys. Rev. Lett \danger{69} (1992) 1849-1851

\bibitem{ct} S. Carlip, C.Teitelboim, Aspects of black hole quantum mechanics and thermodynamics
             in 2+1 dimensions,
             Phys. Rev D. \danger{51} (1995) 622-631, arXiv:gr-qc/9405070


\bibitem{c} S. Carlip, Statistical mechanics of the (2+1)-dimensional black hole, Phys. Rev D
            \danger{51} (1995) 632-637, arXiv:gr-qc/9409052

\bibitem{s}  Zhao Ren, Zhang Sheng-Li,  Canonical entropy of three-dimensional BTZ black hole,
             Phys.Lett. \danger{B}641 (2006) 318-322, arXiv:gr-qc/0608122



\bibitem{kk} K. Krasnov, Black Hole Thermodynamics and Riemann Surfaces, Class. Quantum. Grav.
             \danger{20} (2003) 2235-2250, arXiv:gr-qc/0302073




\bibitem{gi} J.Manuel Garc\'\i a-Islas, Observables in 3-dimensional quantum gravity
             and topological invariants, Class. Quantum. Grav. \danger{21} (2004) 3933-3952,
             arXiv:gr-qc/0401093

\bibitem{bgm} John W.Barrett, J.Manuel Gar\'\i a-Islas, Jo\~ao F.Martins, Observables in the Turaev-Viro
              and Crane-Yetter models, Journal of Mathematical Physics. \danger{48}
              (2007), arXiv:math.QA/0411281



\bibitem{skg} V. Suneeta, R.K. Kaul, T.R. Govindarajan, BTZ Black Hole Entropy from Ponzano-Regge
               Gravity, Mod. Phys. Lett \danger{A14} (1999) 349-358, arXiv:gr-qc/9811071






\bibitem{tv} V.G.Turaev, O.Y.Viro 1992 State Sum Invariants of 3-Manifolds and Quantum
             6j-Symbols, Topology. \danger{31} No 4, 865-902.



\bibitem{jb} John W.Barrett, Geometrical measurements in
              three-dimensional quantum gravity, Int.J.Mod.Phys
              \danger{A18S2} (2003) 97-113, arXiv:gr-qc/0203018


\bibitem{kl} L.H.Kauffman, S.L.Lins, Temperley-Lieb Recoupling Theory and Invariants
            of 3-Manifolds, Princeton University Press. (1994)



\bibitem{d} William Donelly, Entanglement Entropy in Loop Quantum
             Gravity, arXiv:0802.0880v1 [gr-qc]


\end{thebibliography}
\end{document}